# Realizing Edge Marketplaces: Challenges and Opportunities

**Blesson Varghese**
Queen's University Belfast

**Massimo Villari**
University of Messina

**Omer Rana**
Cardiff University

**Philip James**
Newcastle University

**Tejal Shah**
Newcastle University

**Maria Fazio**
University of Messina

**Rajiv Ranjan**
Newcastle University

**Editor:**
Rajiv Ranjan
raj.ranjan@ncl.ac.uk

The edge of the network has the potential to host services for supporting a variety of user applications, ranging in complexity from data preprocessing, image and video rendering, and interactive gaming, to embedded systems in autonomous cars and built environments. However, the computational and data resources over which such services are hosted, and the actors that interact with these services, have an intermittent availability and access profile, introducing significant risk for user applications that must rely on them. This article investigates the development of an *edge marketplace*, which is able to support multiple providers for offering services at the network edge, and to enable demand supply for influencing the operation of such a marketplace. Resilience, cost, and quality of service and experience will subsequently enable such a marketplace to adapt its services over time. This article also describes how distributed-ledger technologies (such as blockchains) provide a promising approach to support the operation of such a marketplace and regulate its behavior (such as the GDPR in Europe) and operation. Two application scenarios provide context for the discussion of how such a marketplace would function and be utilized in practice.

One of the potential business drivers for an *edge marketplace* using Internet of Things (IoT) devices, edge-computing resources, and data science is an enhanced ability to make quicker and better decisions. For example, the reduction in cost and increase in uptake of IoT-based environmental monitoring[1] provides the potential for city managers to augment their decision making both on an operational front using real-time data and analytics and on a strategic front through access to multiple historic observations across multiple sectors.[2]

Enabling marketplaces at the edge and integrating them into existing edge- and cloud-computing infrastructure is nontrivial. Emerging distributed-ledger technologies (DLTs) such as blockchains have promising features that could help in realizing such marketplaces (see Figure 1). However, DLTs will require new extensions before they can be fully leveraged to establish edge marketplaces. Such extensions will require collaboration across multiple disciplines.

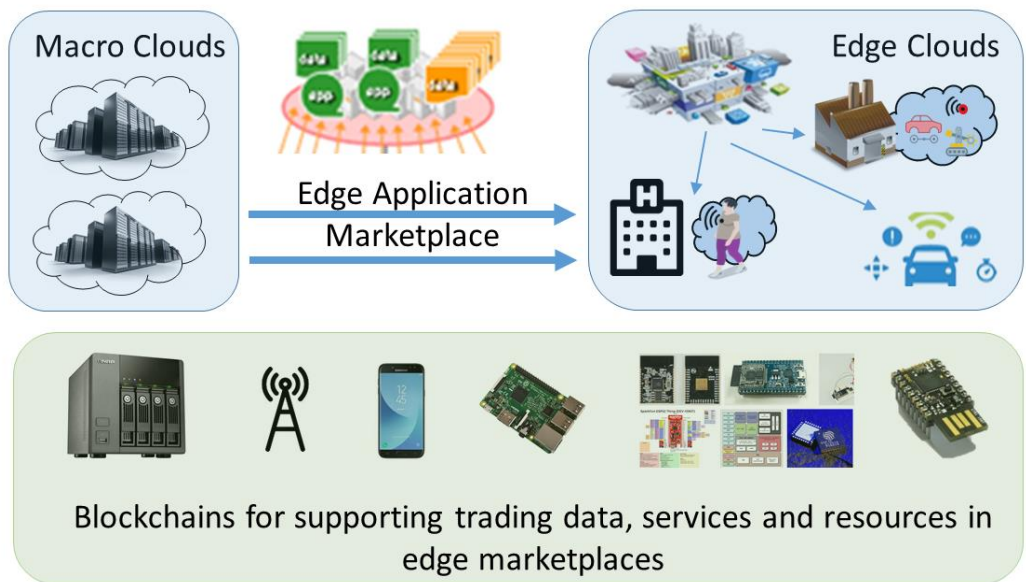

Figure 1. Examples of a smart world driven by the Internet of Things (IoT): from smart homes to smart retail, Industry 4.0, and smart grids.

From a computer science perspective, the existing implementations of DLTs do not support holistic event management and querying, which are required to support trusted and transparent integration of heterogeneous IoT data sources, edge–cloud resources, vendors, and consumers as part of a marketplace. DLT technologies are also intended for use with data that does not change very often, unlike generation rates observed with IoT and edge resources. Ensuring a suitable level of performance in updating the distributed ledger and ensuring consistency remain challenges in this context.

From an economic-modeling perspective, existing cryptocurrencies (a form of digital cash) such as Bitcoin are inappropriate for supporting trading of data, services, and resources in the edge marketplace. From a regulatory perspective, it is not well understood how regulations such as the European General Data Protection Regulation (GDPR) will impact the activities of edge marketplaces with regards to accountability and compliance.

Thus, although DLTs (such as blockchains) have been proposed to provide enhanced security and privacy, understanding the true implication of their use in the context of the GDPR and edge marketplaces remains limited. In this Blue Skies article, we provide insights into these issues by considering two real-world scenarios focusing on smart cities and healthcare.

## SCENARIO 1: SMART CITIES AND DIGITAL TWINS

Cities are complex systems of systems where human, social, technical, infrastructure, and natural systems interact in complex and ever-changing patterns that generate and drive the inevitable trade-offs, compromises, and debates that accompany such decision making. This new paradigm of decision making is enabled partially through high-velocity data streams from in situ environmental monitors (climate, air quality, and traffic flow) and operational data from real-time monitored infrastructure systems (traffic signals, water flow, electrical demand, etc.). This data is coupled with models that enable predictions that are sector specific, such as a high-resolution hydrological flood model, or integrated models across multiple systems (at different levels of granularity).

Data from observations in the real world analyzed stochastically and empirically provides the mathematical foundations and sets the boundary conditions of these models. It is also the means by which the validity of the model outputs can be judged. More recently, observations have been used to parametrize and tune models in real time, providing better solutions driven by real-time observations in the field.[3]

The ability to use data-driven models and high-resolution observational data has been described as a *digital twin*—i.e., a mirror image of a process that is articulated alongside the process in question, usually matching exactly the operation of the process that takes place in real time.[4] A digital twin of a city would provide a means by which planners, engineers, architects, policy makers, etc. can explore decisions and their impacts, governed by observations in the real world.

The development of digital twins for cities or infrastructure systems is predicated on the availability of data. This decision making can take place over hours, minutes, or even seconds, rather than weeks, months, and years. However, such decision making demands larger volumes of (high quality) data at higher temporal and spatial resolutions. Although IoT monitoring is becoming more cost effective, it would be financially impossible for a single entity to meet the capital or operational expenditure of deploying and managing monitoring across a city encompassing all sectors.

A more likely scenario is a pluralistic approach in which sensors and IoT monitoring are deployed by a large number of different bodies for varying operational reasons. Partitioning the monitoring space across multiple organizations such as local authorities, government departments, private companies, infrastructure providers, and even private citizens generates a challenge when trying to create a digital twin of a city encompassing multiple systems.

The above challenge is technical in that there need to be mechanisms to locate, access, query, and share data effectively. Operationally, the maintenance of sensing infrastructure is expensive, and anything that can offset this cost through data sharing would be of enormous benefit. However, this would need an effective marketplace in which the value of this data could be acknowledged and exchanged. One can envisage a scenario in which data collected operationally by a shopping mall on real-time pedestrian flow (velocity, direction, density, etc.), data from public-transportation providers (capacity, passenger numbers, etc.), and data from infrastructure providers (traffic volume, speed, electrical demand, and water flow) would be critical parts of any digital twin of a city.

An edge marketplace for this data provides an ecosystem to offset operational costs and encourage further deployments, and in doing so leads to improved models enhancing the digital twins of systems. Using other people's data in decision making generates questions about data quality, provenance, and trust that require new approaches to ensure the veracity of the data and the maintenance of trust relationships.

## SCENARIO 2: REHABILITATION IN HEALTHCARE

The healthcare scenario is centered on a rehabilitation tool named Lokomat, which is widely used at the IRCCS (Scientific Institute for Research and Healthcare) Hospital in Messina, Italy. Lokomat assists patients in improving their gait and improving the use of their lower limbs according to the physiological gait pattern, and it provides optional partial weight support. Speed, frequency, stride length, and extension of the knee and hip are editable parameters during rehabilitation.

Moreover, patients are involved in an active way. Using virtual reality, patients can drive an avatar in a virtual game. On the basis of mobility conditions, patients may also direct their avatar (right or left) through their hip movements. The LokomatPro version shown in Figure 2 is used. It performs four different gait analyses (assessment tools):

- recording the characteristics of each gait step;
- measuring the isometric force produced by a patient in static conditions;
- measuring hip and knee stiffness during movement, and recording values of relative torque; and
- measuring the patient's passive range of motion without body weight support.

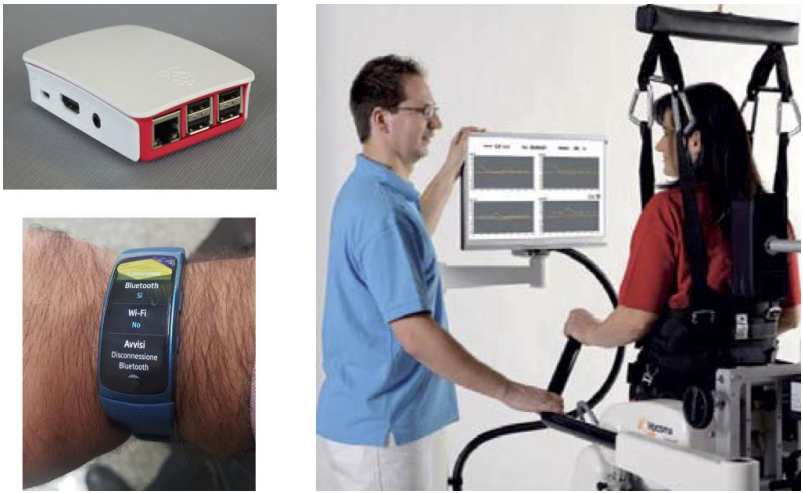

Figure 2. The rehabilitation tool Lokomat, which is used at the IRCCS (Scientific Institute for Research and Healthcare) hospital in Messina, Italy. The tool is connected to a Raspberry Pi 3 (an edge device); Bluetooth/Wi-Fi bracelets allows tracking of practitioners and patients during treatment.

In the context of Lokomat analysis, patients have limited knowledge about how their data is processed and used. For example, the outcome of Lokomat analysis is a comma-separated value (CSV) format file with raw data organized in the header and body portions. The CSV file is stored inside a Raspberry Pi 3 (a type of edge device) that provides a Samba-based shared folder. Additionally, when a patient's data (name, weight, height, time stamp, length of shank and thigh support, etc.) is transferred from one care provider to another, there is often no data integrity verification. All in all, the healthcare data could end up in the wrong hands, at the same time making it difficult to identify fraud.

Hence, similarly to the digital-twin scenario, the healthcare domain needs to implement an edge marketplace for enabling data exchange while preserving data quality, provenance, and trust in a transparent manner.

## THE EDGE

Using the cloud has taught us a variety of things over the past decade.[5] Transferring all our data to the cloud can be expensive and can be a logistical nightmare from a technology standpoint. In addition, data stores still operate as silos, and it can be challenging to transfer data between providers. Not all data needs to be sent to the cloud. The response times between end devices and datacenters over a busy Internet currently are not well suited for latency-critical applications. In the IoT era, more devices and sensors will be connected to the Internet, and we are becoming more concerned about privacy and how data generated at the network edge can be processed.

It is anticipated that at the edge, low-power processors (for example, ARM CPU cores), in contrast to large server processors (for example, Intel Xeon CPUs), will be available.[6] Such processors are available on user devices, such as smartphones or tablets, which can be used to share spare computational resources for facilitating computing on peer devices. There are recent trends of incorporating CPUs on sensors that can be programmed for performing analytics on data that is generated and communicating with higher-level nodes.

Low-power processors could also be made available on a number of edge nodes. For example, existing traffic-routing nodes (such as routers and access points), which are traditionally not employed for general-purpose computing, could be augmented with additional hardware. Alternatively, spare computational resources that might be available on these nodes during off-peak hours could be harnessed. These nodes might be located in public spaces, such as buildings, coffee shops, shopping malls, etc.

In addition, dedicated computational nodes in the form of micro datacenters might be available. These have more computational and storage capabilities than existing or augmented traffic-routing nodes. There are ongoing efforts to make single-board computer clusters such as Raspberry Pi clusters available for utility computing—for example, the Federated RaspberryPi Micro-Infrastructure Testbed project (FRuIT). Specialist hardware, such as accelerators—namely GPUs and field-programmable gate arrays (FPGAs)—can also be incorporated at the edge. ARM Mali GPUs and Nvidia Jetson GPUs are examples of low-power accelerators that can be used at the edge.

It is a significant challenge to incorporate these resources, which will operate in diverse networks and be owned by multiple organizations, into a unified marketplace. Mechanisms for auditing resources will first need to be put in place before a marketplace can even be considered.

## BLOCKCHAINS FOR THE EDGE MARKETPLACE

Blockchains enable data sharing with an immutable recording of all activities performed on the data. Simply put, blockchains provide an unmatched level of proof of integrity of data to ensure auditability and transparency, two crucial requirements for compliance demonstration in the context of edge marketplaces. However, data immutability is in direct conflict with a data subject's "right to erasure" under the GDPR. Thus, although blockchains have been proposed to provide enhanced security, privacy, compliance, and auditability, understanding this technology's implications in the context of the GDPR is paramount before it can be widely adopted for edge marketplaces.

In April 2018, the European Commission announced that European countries will join the European Blockchain Partnership. The aim of the partnership is to exchange expertise and experiences in regulatory and technical domains and to be ready for the beginning of EU-wide blockchain adoption across a digital single market for the benefit of the private and public sectors. The European Commission foresees blockchain technology as the enabler for medical data to be stored and transmitted safely and effectively. Similar thoughts have been echoed by academic researchers.[7,8]

### Public and Private or Federated Blockchain Differences

Blockchains may be broadly classified as public or as private or federated. The public type of blockchain was originally linked to the Bitcoin protocol. In particular, Satoshi Nakamoto in 2008 published "Bitcoin: A Peer-to-Peer Electronic Cash System,"[9] and the first Bitcoin block in the chain got mined in 2009. The Bitcoin protocol is open source; anyone can take the protocol, fork it, modify the code, and start his or her own version of P2P (peer to peer) virtual money.

The main idea of a blockchain is to store unmodifiable transactions on a shared registry (name ledger) thanks to a consensus algorithm among the peers that belong to the distributed ledger. The consensus is achieved in a distributed way among many peers participating in the P2P network. The peers are considered validators of the unmodifiable block in the chain, such that using their private or public asymmetric key and hashing functions over blocks produces unmodifiable pointers to the subsequent new blocks. If the transaction in a block is modified without consensus, then the chain is deemed invalid.

Subsequently, the Ethereum project decided to create its own blockchain, with very different properties from Bitcoin, by introducing *smart contracts*. This concept involves decoupling the application layer from the core blockchain protocol, thereby offering a radical new way to create online markets and programmable transactions. In a public blockchain, there is no need to maintain servers. This naturally reduces the costs of creating and running decentralized applications by leveraging the distributed volunteer P2P network.

Institutions such as banks have understood the benefit of a blockchain as a DLT and created a permissioned blockchain (private or federated). Here, the validator of transactions is not anonymous and public, as in the Bitcoin and Ethereum protocols, but is a member of a consortium or separate legal entities of the same organization. The term "permissioned private ledger" is controversial; therefore, the market refers to this blockchain adoption as a DLT. Distributed and decentralized P2P networks are used as in a public blockchain, but private organizations are involved; hence, these blockchains are referred to as "federated."

Following Gavin Wood's characterization, we have two type of blockchains: permissioned and permissionless (see Table 1).[10]

Table 1. Types of blockchains and their characteristics.[10]

| Blockchain type | Characteristics | | |
|---|---|---|---|
| Permissioned | Faster | Private | Managed |
| Permissionless | Slower | Public | Open |

Figure 3 highlights the main differences between permissioned and permissionless blockchains in public and private or federated blockchain domains. A DLT relies on distributed data storage in which multiple entities hold a copy of the underlying database and all nodes or peers are naturally permitted to participate in achieving consensus, which is the result of arriving at a common truth among nodes. If participation is permissionless, anybody is allowed to participate in the network. If participation is permissioned, participants are selected in advance, and access to the network is restricted to these only. The mode of contribution—permissioned or permissionless—has a different impact on how consensus is reached.

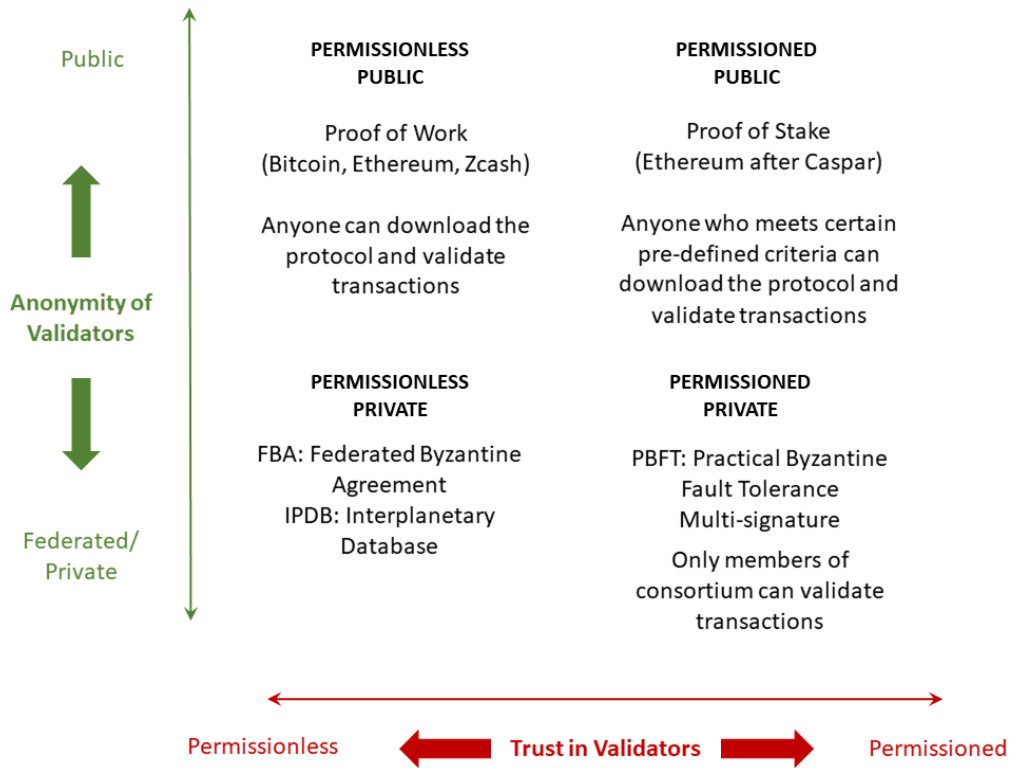

Figure 3. Permissioned versus permissionless blockchains across trust and anonymity axes. Courtesy: https://blockchainhub.net/blockchains-and-distributed-ledger-technologies-in-general/

Fabric and Sawtooth have been both designed toward a similar goal: building a general-purpose, enterprise-level, permissioned blockchain with no a priori defined unique consensus algorithm. As a result, many of the core features of Fabric and Sawtooth are similar, and a supply chain can be built on either of them. Sawtooth works in a variety of languages such as Java, Go, and Ethereum's Solidity smart contracts, whereas Fabric requires the Go programming environment.

The consensus algorithm that Sawtooth offers is Proof of Elapsed Time (PoET), which is hardware assisted, resulting in trivial CPU usage. Sawtooth is good for isolating application logic from the core blockchain logic, making development easier. The consensus protocol of Sawtooth is built on Intel's Secure Guard Extensions (SGX). Sawtooth will probably not see mass adoption in public networks because it relies on Intel chips. However, for private and consortium networks, it is a possible alternative to proof-of-work mining.

## Hybrid Models for Public and Private Blockchains

The original Bitcoin blockchain implementation is affected by a number of important issues. One of the main concerns is its speed, and the consequent scalability of the overall system. To alleviate this, hybrid implementations are reported in the literature—for example, the Lightning Network.[11] These are interesting because they combine the benefit of having a public ledger aimed at virtual currency (a Bitcoin blockchain) along with the scalability and non-anonymity of private or federated blockchains.

# OPEN CHALLENGES

In this article, we consider system- and data-related challenges in using blockchains for creating edge marketplaces. Core security, privacy, and trust-related aspects of blockchains have been reported in a previous Blue Skies article.[12]

## Off-Chain versus On-Chain Storage

As blockchains were originally designed to store relatively small and linear data, it remains unclear how they can cope with complex data storage and querying problems that will arise in the future edge marketplaces. Future research efforts will need to focus on understanding how different combinations of blockchain properties (permissionless, permissioned, and overlay network size) and edge market datatypes impact the security and privacy level of data at rest, in transit, and during processing.

Another issue that will require close investigation is how off-chain versus on-chain storage of market data impacts scalability, security and privacy levels, and the integrity assurance of edge market data. In the former, the data is kept outside the blockchain in an external storage or conventional database, whereas immutable hashes of data indexes are stored on the blockchain. In this configuration, the data can be easily corrected and erased as appropriate. On the other hand, on-chain storage allows the tracking of all changes and versions of edge marketplaces. Although storing market data on a blockchain works well in terms of integrity, security, and privacy, compared to off-chain storage, it suffers from scalability issues.

## Market Knowledge Modeling

A successful edge marketplace must be able to address the core technical challenges of the lack of interoperability and common standards for data use and sharing across different edge service providers. For example, in addition to the GDPR, data exchange involving the UK requires compliance with other national regulations such as the Data Protection Act 2018. Understanding context, data-processing activities, and regulatory requirements in a connected way is fundamental for auditability and compliance.

In the context of an edge marketplace that deals with multiple domains, several ontologies (such as regulatory, data lifecycle, provenance, and sensor ontologies) consisting of concepts from each domain can be used together to provide a comprehensive picture of data exchange over a blockchain infrastructure. Through this integrated knowledge modeling, it will be possible to understand data governance activities, reveal patterns or trends of data management practices, and intelligently analyze the data to reveal the consequences of governance and data management decisions. Furthermore, together with a semantic-reasoning engine, ontologies can be used to query over data to generate actionable insights as well as draw inferences for decision support and automated analyses.

## Provenance

Not all potential providers of data, resources, and applications in an edge marketplace are equal. That inequality might lie in the implicit trust relationship the consumer has with the provider. For example, one might consider data to be from a trustworthy source such as a government agency or regulated body. Similarly, all equipment is not the same and will have different performance levels and characteristics. These parameters might be affected by age or (lack of) maintenance. In all these cases, the ability to drill down from the consumer to the provider and understand the provenance of the data being used and ancillary operational metadata is key to providing a trustworthy and value-driven data economy and edge marketplace.

Another compelling growing need for developing provenance of blockchain infrastructure is related to trust and transparency. For example, consider trading the data gathered from a driver's driving habits for exchange with vehicle insurance providers. In addition to a user's willingness for data to be shared, each individual edge marketplace ought to be governed by principles that allow other marketplaces to trust them. For example, is the same data of the user provided to all insurance providers, or are some providers preferred over others in the marketplace? A provenance framework that enforces transparency of data in this regard is essential and will need to be implemented.

Additionally, which actor within the marketplace is responsible if there is a misuse of user data will need to be articulated. For this, all trading points within a market will need to be identified, and contracts, such as service-level agreements that bind the actors at these points, will need to be formulated. This is not an easy task, especially when the data footprint is widespread across many intermittent and heterogeneous edge resources. For example, what happens if an edge resource is used to process data if there is a breach but subsequently is not available anymore as a resource in a marketplace? Also, what auditing mechanism should be in place in this case?

Hence, it will be necessary to develop new techniques that enable scalable querying and auditing of events over a blockchain, to verify the type of operation (the data, resources, and application involved) and the time at which it was carried out.

## Federated Edge Markets

One benefit of using blockchains at the edge is sharing data in a transparent and auditable manner. This characteristic can be leveraged to share data across different edge markets so that the marketplace can be federated.

For example, consider trading data between cities for improving their digital twins or trading the data gathered from a driver's driving habits that is exchanged with vehicle insurance providers. The associated challenges and questions arising are numerous.

The first challenge is a user's privacy concerns. The key question that will need to be addressed is, how comfortable would a user be in exposing data he or she has generated to third parties? This data could be sensitive or might inherently be used to penalize a user. For example, a driver might have to pay a higher premium for insurance should his or her driving habits be passed on to insurance providers. This raises a further question: How can a federated marketplace be GDPR compliant should a user wish to exercise his or her right to erasure?

A second challenge is regulatory compliance. The edge marketplaces could be operating across different geographic locations and industries where the law of the land and the regulatory bodies are vastly different. In this context, it is essential for each market to know whom it is trading with; in addition, each federated marketplace might require its own regulatory body to ensure that individual markets are compliant.

A third challenge is related to the implementation of the technology. Currently, we do not know the shape and form that interfaces between marketplaces should take. These interfaces will need to provide different levels of abstraction of data based on which markets they interact with. For example, what information should be concealed and exposed between marketplaces? The key question that will need to be addressed is, which entity will manage data at this interface? Although a federated marketplace has obvious advantages for sharing data, the open challenges in this area require significant research efforts from academic institutions, government bodies, and the private sector.

To summarize, all of the above requirements necessitate development of geo-blockchains that can scale across multiple marketplaces, across multiple administrative jurisdictions, and possibly across multiple countries that implement different types of governance and compliance standards.

## Market Models

An edge marketplace that enables multiple providers to coexist could be beneficial from a number of different perspectives:

- It would improve access to potential services for data processing, enabling a user to carry out analysis closer to the point of data generation.
- It would enable enhancement of capabilities directly accessible (or owned) by a user without the need to use a cloud platform.

In "Incentivising Resource Sharing in Edge Computing Applications,"[13] we consider a number of potential revenue models for supporting market actors at the network edge, such as the following.

One model is *pre-agreed edge resource contracts*. Using this approach, a user application would rely on informative, detailed contracts that adequately capture the circumstances and criteria that influence the performance of the externally provisioned services. A user application proxy is expected to have pre-agreed contracts with specific edge resource operators and would interact with them preferentially. This also reduces the potential risks incurred by the user in utilizing such services. In performance-based contracts, a user application would need to provide a minimum level of performance (e.g., availability), which would be reflected in the associated price.

Consider the following scenario to illustrate this business model. A coffee chain offers contracts for the use of edge resources operated by it across a city or country. A user wishing to use these resources would need to agree to have

- a security certificate provided by this coffee chain and
- a pre-agreed subscription for the use of these resources.

With an increasing number of branches or locations of this coffee chain, a user would have a greater choice of locations available for use. This is equivalent to accessing wireless networks (Wi-Fi) at locations offered by a particular provider. Such a coffee shop chain might also decide to enter into preferential agreements with cloud datacenter operators (e.g., public-cloud providers) to integrate its regional edge resources with datacenters operated by the cloud provider.

Another model is *edge resource federation*. In this model, multiple operators can collaborate to share workloads within a particular area, and have preferred costs for the exchange of such workloads over their edge resources. This is equivalent to alliances established between airline operators to serve particular routes. To support such a federation, security credentials between operators of edge resources must be pre-agreed. This is equivalent to an extension of the pre-agreed edge-resource-contracts business model, in which edge resource providers across multiple coffee shop chains can be federated, offering greater potential choice for a user.

A third model is *edge–cloud exchange*. In this model, a user application would initially be scheduled to run on a cloud datacenter in the first instance. If the datacenter has congestion or if the network connecting the user application to the datacenter is intermittent or unusable, a cloud provider could then outsource computation to an edge resource if it is unable to meet the required quality-of-service targets (e.g., latency). A cloud resource provider could use any of the approaches outlined above—i.e., dynamic discovery, preferred resources, or the choice of an edge within a particular group. A cloud datacenter operator needs to consider whether outsourcing could still be profitable given the type of workload a user device is generating.

Devising tools and techniques to support each of the above market models over blockchain infrastructure presents unique research challenges.

## CONCLUSION

The emerging availability and capability of devices at the edge of the network provide new opportunities for application developers. Whereas centralized datacenter-based approaches have gained significant traction in recent years, there is now the realization that migrating large data volumes (or computational jobs) for analysis to a central location can be costly, and in some instances unnecessary. Creating a computing system that integrates resources across both the network edge and cloud datacenters is essential for latency sensitive, stream-based applications.

We described two scenarios—smart cities and healthcare—that have these characteristics. Given this context, we described a marketplace for services that can exist at the network edge. Such a marketplace would enable multiple providers to offer services, driven by requirements of user demand, security, and availability. An edge marketplace can be supported through DLTs (such as blockchains) to ensure that data about services and their engagement in particular transactions can be kept in a tamper-proof record. This enables the establishment of trust for particular providers, ensuring that a user is able to get accurate information about previous interactions.

We also described mechanisms for realizing a practical edge marketplace, enabling exchange of data and services. Understanding revenue and incentive models for realizing an edge marketplace remain important challenges.

## ABOUT THE AUTHORS

**Blesson Varghese** is a lecturer in computer science at Queen's University Belfast. His research interests include next-generation distributed systems that leverage the cloud–edge continuum. Varghese received a PhD in computer science from the University of Reading. Contact him at b.varghese@qub.ac.uk.

**Massimo Villari** an associate professor of computer science at the University of Messina. His research interests include cloud computing, the Internet of Things, big data analytics, and security systems. Villari received a PhD in computer engineering from the University of Messina. Contact him at mvillari@unime.it.

**Omer Rana** is a full professor of performance engineering at Cardiff University's School of Computer Science and Informatics. His research interests include performance modelling, simulation, the Internet of Things, and edge analytics. Rana received a PhD in computing from Imperial College London. Contact him at ranaof@cardiff.ac.uk.

**Philip James** is a senior lecturer at Newcastle University's School of Civil Engineering and Geosciences. His research interests include the Internet of Things, next-generation analytics, and spatial-data management. James received a BA in Japanese from Newcastle University. Contact him at philip.james@ncl.ac.uk.

**Tejal Shah** is a research associate in the Data to Knowledge (D2K) research group working primarily on the Connected Health Cities program at Newcastle University. She's an expert on the research, development, and application of Semantic Web and knowledge representation technologies such as ontologies and vocabularies to address complex information management issues. Shah received a PhD in Computer Science and Engineering from the University of New South Wales. Contact her at tejal.shah@ncl.ac.uk.

**Maria Fazio** is an assistant researcher in computer science at the University of Messina. Her main research interests include distributed systems and wireless communications. Fazio received a PhD in advanced technologies for information engineering from the University of Messina. Contact her at mfazio@unime.it.

**Rajiv Ranjan** is a chair professor of computing science and the Internet of Things at Newcastle University. His research interests include the Internet of Things and big data analytics. Ranjan received a PhD in computer science and software engineering from the University of Melbourne. Contact him at raj.ranjan@ncl.ac.uk.